# Metal Fillers as Potential Low Cost Countermeasure against Optical Fault Injection Attacks

Dmytro Petryk[1], Zoya Dyka[1], Jens Katzer[1] and Peter Langendörfer[1,2]

[1]*IHP – Leibniz-Institut für innovative Mikroelektronik*
*Frankfurt (Oder), Germany*

[2]*BTU Cottbus-Senftenberg*
*Cottbus, Germany*

*Abstract*—Physically accessible devices such as sensor nodes in Wireless Sensor Networks or "smart" devices in the Internet of Things have to be resistant to a broad spectrum of physical attacks, for example to Side Channel Analysis and to Fault Injection attacks. In this work we concentrate on the vulnerability of ASICs to precise optical Fault Injection attacks. Here we propose to use metal fillers as potential low-cost countermeasure that may be effective against a broad spectrum of physical attacks. In our future work we plan to evaluate different methods of metal fillers placement, to select an effective one and to integrate it as additional design rules into automated design flows.

## I. Motivation

Wireless sensor networks (WSN) are more and more used in automation systems and in the area of critical infrastructure protection. One of the requirements for such devices is to keep the processed and transmitted data confidential and to ensure their integrity. This can be achieved by applying cryptographic algorithms.

The cryptographic strengths of a cipher algorithm depends according to the definition of Kerckhoff only on the used cryptographic key that is kept secret [1]. This means a potential attacker may know the algorithm itself, the plain text, the encrypted text and even the length of the key. In such a situation the attacker can test different numbers in order to reveal the key. The cryptographic keys have to be sufficiently long so that the time for brute forcing is sufficiently long. The situation changes dramatically if the attacker knows not only the input and output values but also intermediate values or physical parameters such as energy consumption and its distribution during the execution of the operation. Temperature, electromagnetic emission and other physically measureable parameters are a kind of "side effects". If the device is physically accessible the attacker can reveal the private/secret key analyzing side effects measured. These attacks are side channel analysis (SCA) attacks.

Another kind of powerful attacks are fault injection (FI) attacks. In these attacks faults are induced into an ASIC, e.g. in order to get access to internal data. FI attacks can be performed by various sources of faults: voltage, temperature, electromagnetic radiation, etc. The purpose of FI attacks is to inject a fault that switches the device into an erroneous operation mode. Exploitation of such an erroneous operation mode and monitoring its output may leak the secure data. In this work we discuss FI attacks performed with a laser, i.e. we concentrated on localized optical FI attacks. Optical FI attacks are feasible due to the internal photoelectric effect. This effect is based on interaction of silicon with laser light. Details about the internal photoelectric effect can be found in [2], [3].

Design and implementation of crypto hardware that is resilient against fault attacks is an extremely sophisticated task. At least, currently, there are no guidelines how to do it. The core idea discussed in this paper is to use metal fillers to prevent manipulation of devices by laser-based FI attacks.

The paper is structured as follows. Section II briefly describes the IHP technologies. Section III describes the optical FI setup used to perform the attacks described here. Section IV present a short description of the attacked chips and the obtained results. Section V discusses metal fillers as low cost countermeasure against physical attacks and compares it with existing radiation hardening techniques. Section VI concludes this work.

## II. The IHP CMOS technologies

In order to prepare precise laser FI attacks knowledge about the internal structure of the chip is necessary. We use the IHP CMOS technology [4] as an example for our experiments. The IHP 8 inch wafer fab for research and production can manufacture chips in a 250 nm and in a 130 nm technology. In this section we give some details about these technologies. The knowledge of these details can be used not only by attackers for attack preparation but also by designers as effective countermeasures.

The thickness of the substrate of the 8 inch wafers is about 0.7 mm. It is usually thinned to a thickness of about 0.2-0.3 mm chips before they are used in devices. **Fig. 1** shows a cross-section of a chip in the IHP 250 nm CMOS technology schematically, i.e. wires in different metal layers and their interconnectors are zoomed in to illustrate the physical size of the chip structure.

Chips produced in the IHP 250 nm technology consist of 5 metal layers: 3 thin and 2 thick metal layers. The interconnectors between metal layers are called vias. The bottom metal layer – metal layer 1 – is usually reserved for connecting transistors to power supply. In other metal layers the connection between gates is realized. The wires within a metal layer are (usually) parallel to each other while wires in neighboring layers are orthogonal to each other.

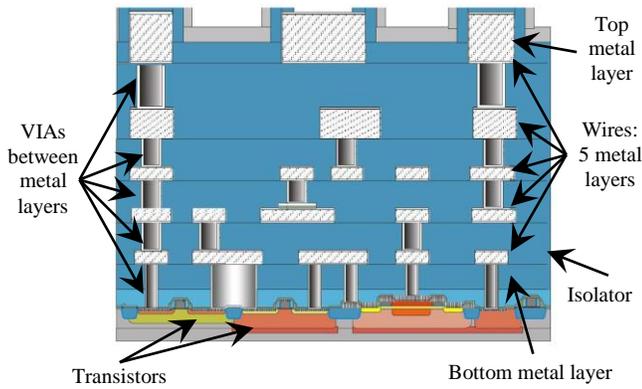

**Fig. 1.** Schematical cross-section of a chip on the example of the IHP 250 nm CMOS technology.

Chips produced in the IHP 130 nm technology consist of 7 metal layers: 5 thin and 2 thick metal layers. Due to technology requirements both IHP technologies, i.e. the 250 nm and the 130 nm technology have metal fillers.

Metal fillers are small metal structures – rectangles – placed in different metal layers. Metal fillers are applied as standard means to increase the mechanical stiffness of wafers during manufacturing process. If the global metal density of the metal layer does not meet the technology requirements it is filled with metal fillers. They are placed between the wires in each metal layer if required. The placement of the metal fillers is a mandatory step of the layout process that is performed automatically.

### III. SETUP FOR OPTICAL FI ATTACKS AT IHP

To perform laser based FI attacks we used a setup available at the IHP, see **Fig. 2**. It contains: a 1st generation Riscure Diode Laser Station (DLS) [5] placed in a safety box, a stable power supply, an oscilloscope and a PC with dedicated FI software.

The DLS consists of: a laser source, a spot size reducer, a microscope camera, a source of light for target illumination, a DLS body, an optical system and a high-precision X-Y positioning stage [6]. The DLS is equipped with two multi-mode laser sources. In 1st generation of the Riscure DLS only one laser source can be used simultaneously. According to [5] this DLS has following specifications:

- multi-mode laser sources:
    - red (808 nm), maximum power is 14 W;
    - infrared (1064 nm), maximum power is 20 W;
    - pulse duration in a range of 20 ns – 100 μs;
    - elliptical spot sizes of 60×14 μm$^2$, 15×3.5 μm$^2$, 6×1.5 μm$^2$ or 3×0.8 μm$^2$;

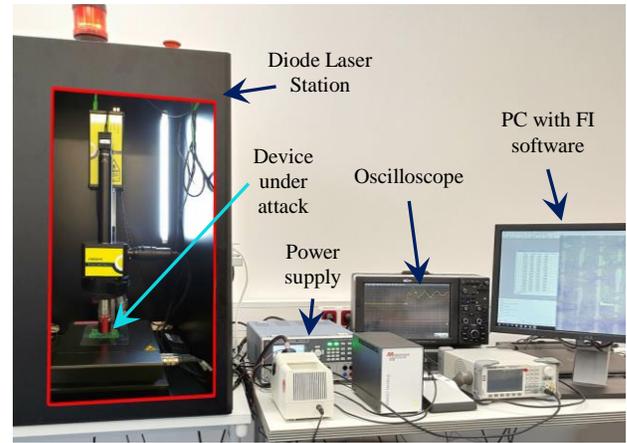

**Fig. 2.** Optical fault injection setup available at the IHP.

    - filter: 0.1 %, 1 %, 10 %;
- single-mode laser source [7]:
    - red (808 nm), maximum power is 0.848 W;
    - pulse duration in a range of 2 ns – Continuous Wave (CW);
    - circular spot sizes of 15 μm$^2$, 4 μm$^2$, 1.5 μm$^2$ or 1 μm$^2$;
- magnification objectives: 5×, 20×, 50×, 100×;
- X-Y table with 3 μm accuracy and 0.05 μm [5] step size.

We applied the red multi-mode laser source in all our experiments described here. Hence, all attacks have been performed through the front-side of the tested chips. Additional details about the optical FI setup can be found in [3].

### IV. EFFECTIVENESS OF PERFORMED FI ATTACKS

#### A. Attacks against the IHP CMOS 250 nm technology

Our first device under attack (DUA) is IHP's "Libval025" chip manufactured in the IHP 250 nm technology with 5 metal layers. Originally, the chip was designed to measure signal propagation delays through different types of gates: invertors (INV), NAND gates, NOR gates and flip-flops (FF). Each libval-structure consists of 16 small independent circuits. Each circuit is a sequence of a single type of gates, e.g. a sequence of AND gates, or invertors only.

The structural scheme of Libval025 is shown in **Fig. 3**.

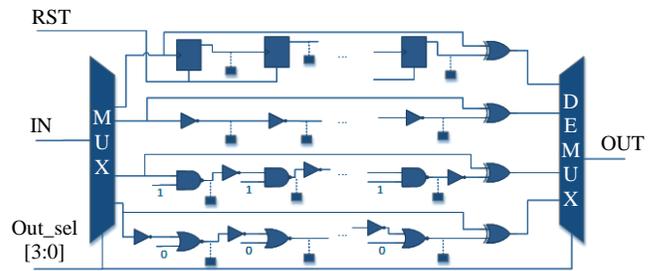

**Fig. 3.** Structural scheme of Libval025.

**Fig. 4** shows Libval025 chips bonded on a PCB.

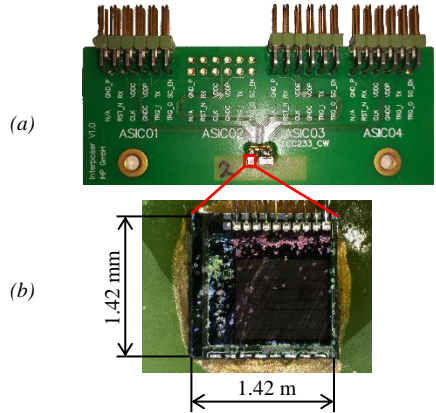

**Fig. 4.** The 3 bonded Libval025 chips on a PCB *(a)* and a single Libval025 chip zoomed in *(b)*.

We attacked all 3 Libval025 chips in our experiments.

The Libval025 chips described here were designed and produced about 20 years ago. We selected this chip for the experiments due to the fact that it was produced without metal fillers yet. Since Libval025 has no metal fillers the internal structure of the chip is clearly visible, e.g. through a microscope camera. **Fig. 5** shows the surface of a Libval025 chip without metal fillers captured using microscope camera with 100× magnification objective.

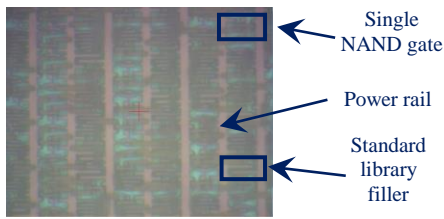

**Fig. 5.** Surface of a Libval025, captured using a 100× magnification objective.

In the Libval025 chips, single gates can be easily localized by optical inspection using a microscope camera or even a conventional microscope. Hence, optical FI attacks can be performed fast and effectively. Due to the parameters of our laser equipment and the size of the attacked gates, a single gate can be selected and attacked, i.e. we performed a localized FI attack. The single selected gate can be, e.g. a flip-flop of a register that can contain the secret/private key.

Results of front-side FI attacks on the Libval025 chips show that faults can be injected successfully in all 4 types of gates, i.e. FF, NAND, INV and NOR. The faults are repeatable for all 3 tested chips with a slight deviation of the applied laser beam parameters, i.e. intensity and/or pulse duration. Both transient and permanent faults were successfully injected. The latter was achieved by a significant increase of the laser beam power that subsequently led to the damage of the internal structure. A detailed description of the experiments done with the Libval025 chips can be found in [2]. A short summary of the attack results is given in TABLE I, see section *IV-D*.

### B. Attacks against the IHP CMOS 130 nm technology

Next we attacked the Libval chip manufactured in the IHP 130 nm technology. We denote this chip as "Libval013" in the rest of this paper. The structure of Libval013 is the same as for Libval025, i.e. the chip contains 4 types of gates: INV, NOR, NAND and FF. Opposite to Libval025 the attacked Libval013 has metal fillers. They are placed in different metal layers. Due to the metal fillers the internal structure of Libval013 is not visible with a microscope camera. **Fig. 6** shows the surface of Libval013 captured by microscope camera using a 5× magnification objective and a part of Libval013 surface zoomed in using a 100× magnification objective.

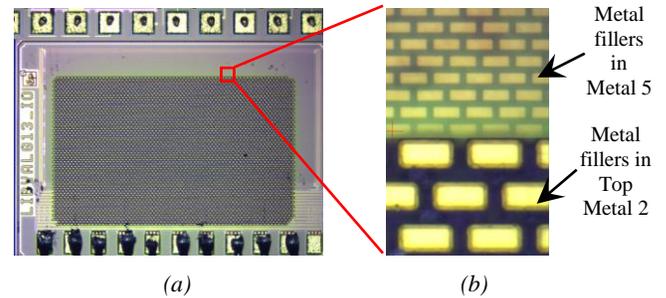

**Fig. 6.** Surface of a Libval013 captured using a 5× magnification objective *(a)* and a part of Libval013 surface zoomed in using a 100× magnification objective *(b)*.

The metal fillers in Libval013 have different sizes in different metal layers and can be placed on top of each other. The metal fillers in the Top Metal 2 are the biggest but the distance between the fillers – the "gap" – in this layer is also the biggest one (see **Fig. 6**–(*b*)). Hence, we expected that in our experiments more successful FI will be observed when attacking the gates "placed" under the "big gaps" of large metal fillers.

The results of the front-side optical FI attacks on Libval013 confirmed our assumption. It was possible to inject faults in all type of gates. However the area of the chip that is sensitive to laser irradiation is reduced compared to Libval25. The state of the gates covered with metal fillers was not influenced in our laser experiments. Transient faults were successfully injected only in gates that are not covered with metal fillers. No damage of the internal structure was observed even if we illuminated the fillers with the maximum red laser beam output power over a relative long time (100 µs). The overall success rate of FI attacks is significantly reduced compared to Libval025.

### C. Attacks against IHP RRAM structures

Additionally, we attacked the IHP Resistive Random Access Memory (RRAM) chips. The 4 kBit RRAM chips were manufactured in the IHP 250 nm technology. A single RRAM cell in IHP chips is based on a 1 transistor – 1 resistor (1T-1R) architecture. The memory element in the IHP RRAM cell is composed of a Metal Insulator Metal (MIM) stack. **Fig. 7** shows a Transmission Electron Microscope (TEM) image of an IHP RRAM cell based on the 1T-1R architecture.

The MIM structure is placed on top of Metal 2 and connected with Metal 3 through a tungsten via. The MIM structure is of interest here since we attacked a standalone RRAM chip, i.e. the NMOS transistor cannot be switched by laser irradiation.

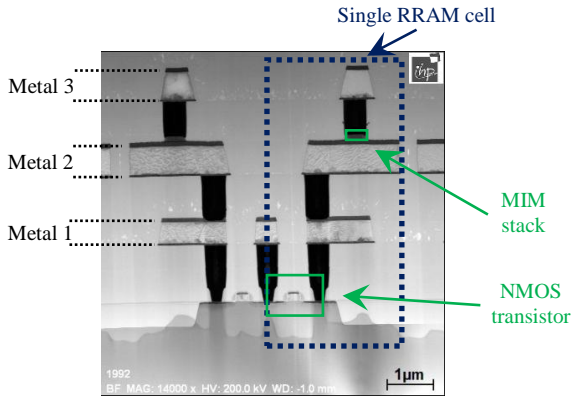

**Fig. 7.** TEM image of 1T-1R IHP RRAM cell, taken at IHP.

Additional details about IHP RRAM structures, i.e. switching behavior, purpose of transistor and MIM structure, can be found in [3], [8]. The IHP 4 kbit RRAM chips have metal fillers only in two metal layers, i.e. in Top Metal 2 and Top Metal 1 [9]. The size of all RRAM cells is the same. Thus, the placement of the cells as well as the one of metal fillers is periodical and the structure looks very regular. Due to the metal fillers the arrangement of the RRAM cells is not visible through the microscope camera. **Fig. 8** shows the attacked 4 kbit RRAM chip, the part of its surface was captured using a 100× magnification objective and a cross section image of the chip that was made with a Scanning Electron Microscope (SEM). The chip was prepared for the SEM-imaging using a Focus Ion Beam (FIB) cut in an IHP laboratory.

The front-side FI attacks on the RRAM chip show that it is possible to induce both transient and permanent faults into the chip. The latter is however achieved with significantly higher laser beam parameters than for Libval025. Analysis of obtained data shows that success of optical FI attacks depends on the location of metal fillers atop the RRAM cell. The metal fillers in the RRAM chip have different thickness but similar width and length. They are placed in Top Metal 1 and Top Metal 2, sometimes exactly under each other. Hence, areas that are not covered by metal fillers in Top Metal 2, can also not be covered by metal fillers in Top Metal 2. Thus, leaving "gaps" the laser beam can freely go through and illuminate the cell. In our experiments the RRAM cells placed under "gaps" were successfully influenced with the laser beam. Nevertheless some of the RRAM cells that are covered by metal fillers were also influenced but the success rate of the FI attacks for these cells is significantly lower. A detailed description of the experiments done with IHP RRAM chips and results of the FI attacks performed can be found in [3].

*D. Attack results summary*

TABLE I summarizes the results of the front-side optical FI attacks for different IHP chips.

The criterion of success of optical FI attacks was determined as follows ($N$ is the percentage of successfully attacked gates/cells from all attacked gates/cells):

- very high: $(90 \leq N \leq 100)$ %;
- high: $(50 \leq N < 90)$ %;

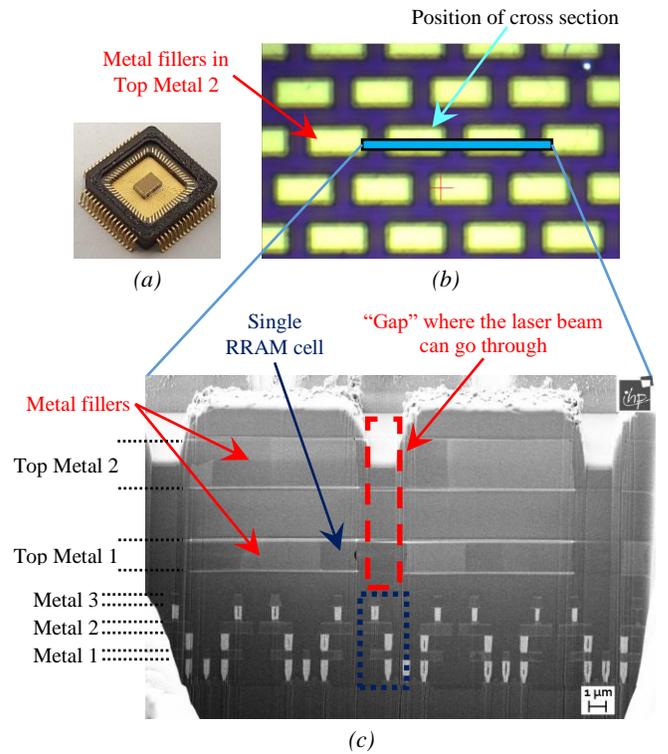

**Fig. 8.** Attacked RRAM chip *(a)*; a part of the chip surface, captured using 100× magnification objective *(b)*; SEM cross section image (FIB cut) of the attacked IHP RRAM chip *(c)*.

TABLE I. SUCCESS OF OPTICAL FI ATTACKS FOR DIFFERENT IHP CHIPS

|  | Device under attack | | | |
|---|---|---|---|---|
|  | Libval025 | Libval013 | | RRAM |
| Metal fillers, placement | no fillers | placed in Metal 5 | placed in Metal 5 and Top Metal 1, and Top Metal 2 | "gaps" between metal fillers | placed in Top Metal 1 and Top Metal 2 |
| Success of FI attacks | very high | low | very low | moderate | low |

- moderate: $(25 \leq N < 50)$ %;
- low: $(10 \leq N < 25)$ %;
- very low: $(0 \leq N < 10)$ %.

V. METAL FILLERS AS LOW COST COUNTERMEASURE

Metal fillers and connecting wires are obstacles for laser beam propagation since they absorb/reflect the laser light making harder it for a laser beam to reach the attacked gate. The results of our experiments given in TABLE I confirm the fact that metal fillers reduce the success of front-side optical FI attacks significantly.

The idea to prevent optical FI attacks as well as localized electromagnetic analysis attacks using metal obstacles for a laser beam propagation is not new. For example in [12]-[14] it was proposed to use the metallization layers for supplying ASICs with VDD and GND as a countermeasure for semi-invasive front-side attacks. Both supply voltages, implemented as metal planes, can be placed for example in top metal layers to prevent optical access to the transistor level while the device is fully functional. Additionally, it may be effective against microprobing, too.

Alternatively, efficient countermeasures can be designed to prevent/mitigate radiation influence using redundancy. These are hardware Triple Modular Redundancy (TMR) [10], Junction Isolated Common Gates (JICG) [11], Dual Interlocked Storage Cell (DICE) [12], etc. They have been studied intensely and already experimentally proved their fault-tolerance to radiation influence [10], [11], [12]. The resistance of the radiation hardened gates against manipulation as well their resistance against side channel analysis attacks have to be investigated. The need of this investigation is discussed in [15]. However implementation of such countermeasures usually requires hardware redundancy, e.g. triplication in TMR [10], doubling in JICG [11], doubling or triplication in DICE cells [12]. Due to this fact, such countermeasures require increased area on a silicon die compared to non-radiation hardened implementations. In order to increase the resistance of the redundant elements against laser fault injection the elements cannot be placed next to each other. For example, if TMR flip-flops will be placed close to each other a laser beam with a relatively large spot size, e.g. the size of two flip-flops, can influence them simultaneously with a high probability. Thus, placement of such redundant elements and arrangement of connection wires between them have to be considered. This usually leads not only to increased area but also to a modification of the automated standard design flow. On the other hand, since implementing these countermeasures requires additional active elements, the device's power consumption increases. Simultaneously, the increasing number of elements can cause a degradation of the performance. Thus, designers have to find a compromise between device performance, fault tolerance and resistance against SCA attacks.

Currently, metal fillers are a technology requirement of IHP that cannot be excluded from design flow. We propose to use the metal fillers as a kind of low-cost but effective countermeasure. To counter optical FI attacks efficiently, the placement of metal fillers has to be carefully evaluated. For example, default placement rules for metal fillers in the automatic design flow does not guarantee efficient mitigation against optical FI attacks since their coverage is not dense enough, i.e. a lot of cells are still not covered, e.g. in IHP RRAM chip. Thus, design rules with respect to metal fillers placement as well as their size and shape have to be reconsidered, i.e. a modification of the automated design flow is required. To comply with this task the areas that are sensitive to laser irradiation have to be determined for each type of gates. After it, the metal fillers can be placed so that all sensitive gate areas will be covered. Subsequently this "intelligent" placement of metal fillers methodology can be automated and implemented in the design flow.

In comparison to the countermeasures mentioned at the beginning of this section the metal fillers have several advantages. It is expected (but still has to be proven) that the intelligent placement of metal fillers will not cause a big overhead in chip area since it does not require doubling or triplication of elements, as metal fillers are more or less only arranged differently in the respective metal layers. The other advantage is that the metal fillers do not consume any power, i.e. the overall power consumption of the device does not increase. Hence, applying metal fillers and automatization of their intelligent placement can be a highly attractive, practical and low-cost countermeasure against optical inspection of chips, optical/laser FI attacks and – eventually even – localized electromagnetic analysis attacks. So, metal fillers can be a low-cost effective countermeasure against a broad spectrum of attacks. In our future work we plan to consider various methods of metal fillers placement in order to find the solution that successfully mitigates/prevents most of optical FI attacks.

In our future work we plan to consider various methods of metal fillers placement in order to find the solution that successfully mitigates/prevents most of optical FI attacks.

## VI. CONCLUSION

In this work we discussed the results of our precise localized optical fault injection attacks. Our experimental results confirm the fact that the metal fillers placed in different metal layers of an ASIC can significantly influence the success of front-side optical FI attacks. Since the implementation of metal fillers is currently a requirement of IHP technology they can be used as a low cost countermeasure/mitigation technique if they cover the gate areas that are sensitive to optical FI attacks. In our future work we plan to investigate the sensitivity of the gates to laser irradiation. Our goal is to design a methodology for "intelligent" placement of metal fillers. The methodology has to be automated and integrated into the design flow.

ACKNOWLEDGMENT

This project has received funding from the European Union's Horizon 2020 research and innovation program under the Marie Skłodowska-Curie grant agreement No 722325.